\documentstyle[epsfig]{mn2e}

\title[The eclipsing Intermediate Polar V597 Pup]
{The eclipsing Intermediate Polar V597 Pup (Nova Puppis 2007)}
\author[Brian Warner and Patrick A.~Woudt]
       {Brian Warner\thanks{email: Brian.Warner@uct.ac.za},
Patrick A.~Woudt\thanks{email: Patrick.Woudt@uct.ac.za}\\
        Department of Astronomy, University of Cape Town, Private Bag X3,
        Rondebosch 7701, South Africa}
\date{2009 April 11}

\begin{document}

\maketitle

\begin{abstract}
Photometric observations of V597 Pup made in 2008, 9.1 mag below maximum, 
4 months after eruption, showed no certain orbital modulation but exhibited 
a quintuplet of oscillations centred on a period 261.9 s and uniform splitting 
at a frequency $\sim$ 2.68 h$^{-1}$. One year later the system had fallen in 
brightness by a further 2.5 mag, showed deep eclipses with a period of 
2.6687 h, and the 261.9 s modulation at a reduced amplitude. There is often 
power near the `subharmonic' at 524 s, showing that the shorter periods 
observed are actually first harmonics.

V597 Pup is thus an intermediate polar and is in the `orbital period gap'. 
Furthermore it is the first to show a prominent secondary eclipse, caused by 
passage of the optically thick disc in front of the irradiated side of the 
secondary star. 
\end{abstract}

\begin{keywords}
binaries -- close -- novae -- stars: oscillations --
stars: individual: V597 Pup, cataclysmic variables
\end{keywords}

\section{Introduction}

\begin{figure}
\centerline{\hbox{\psfig{figure=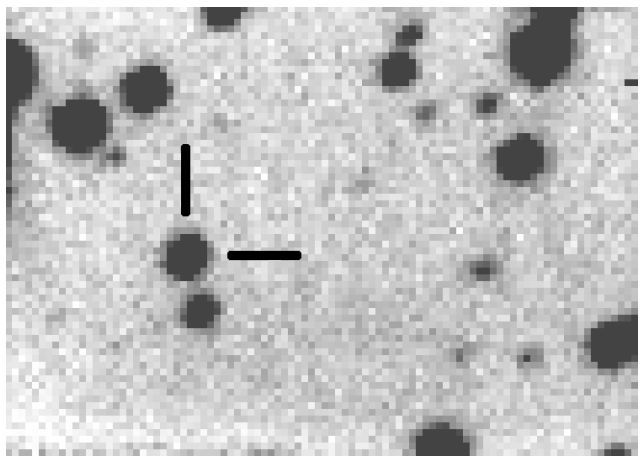,width=8.4cm}}}
  \caption{CCD image of V597 Pup (indicated by the markers) taken on
2009 March 18 (out of eclipse). The field of view is 50 by 34 arcsec, north
is up and east is to the left.}
 \label{warnerfig1}
\end{figure}

V597 Pup was discovered as a nova at V $\sim$ 7.5 on 14 November 2007 
(Pereira, McGaha \& Rhoades 2007), reaching V $\sim$ 6.2 at maximum and 
declining smoothly for at least 5 magnitudes as a very fast nova, with 
$t_2$ = 2.5 d (Naik, Banerjee \& Nashok 2009), which is nearly the shortest 
recorded (only V838 Her and MU Ser are shorter: Table 5.2 of Warner 1995). 
Its infrared red spectra and X-Ray emission (as a super-soft source) in 
early stages of decline are discussed by Naik et al., who classify it as 
a He/N nova and deduce that it has a white dwarf mass close to the 
Chandrasekhar limit.. By March 2008 it had reached V $\sim$ 15.3; our later 
observations place it (out of eclipse) at $\sim$ 17.7 in February 2009 and 
$\sim$ 17.9 a month later. No post-eruption spectra have been published.  
A possible pre-eruption image at V $\sim$ 20 is present on the Digitized 
Sky Survey (Pereira et al.~2007). A finding chart prepared from one of 
our CCD images is given in Fig.~\ref{warnerfig1}. 
The position of V597 Pup appears to coincide with the proposed 
pre-eruption image. 

In Section 2 we give the results and an initial analysis of our observations. 
In Section 3 we look at the overall geometry of V597 Pup. Section 4 
expands on the analysis by treating V597 Pup as an eclipsing intermediate 
polar, and Section 5 summarizes our interpretation of this interesting 
cataclysmic variable. 

\section{Observations}

We used the University of Cape Town (UCT) CCD photometer, as described by 
O'Donoghue (1995), in frame transfer mode and with white light, on the 
1.9-m (74-in)  reflector at the Sutherland site of the South African 
Astronomical Observatory (SAAO). Tab.~\ref{warnertab1} contains a detailed 
observing log. Our magnitude scale was derived using hot white dwarf 
standards, but because of the non-standard spectral distributions of the 
spectrum of the nova remnant, and the use of white light, our magnitudes 
approximate a V scale only to $\sim$ 0.1 mag. 

\begin{table}
 \centering
  \caption{Observing log of photometric observations}
   \begin{tabular}{@{}llccrl@{}}
 Run      & Date of obs.          & HJD of first obs. & Length  & t$_{in}$  &  V \\
          & (start of night)      &  (+2450000.0)     & (h)     & (s)      & (mag) \\[10pt]
S7807 & 2008 Mar 15 & 4541.26125 & 2.77 &  6 & 15.2   \\
S7808 & 2008 Mar 16 & 4542.24472 & 3.53 &  6 & 15.2   \\
S7810 & 2008 Mar 17 & 4543.27429 & 3.82 &  6 & 15.3   \\
S7812 & 2008 Mar 18 & 4544.25449 & 3.30 &  6 & 15.3   \\
S7828 & 2009 Feb 25 & 4888.26865 & 7.30 & 30 & 17.8$^\dag$  \\
S7830 & 2009 Feb 26 & 4889.28184 & 6.95 & 30 & 17.7$^\dag$  \\
S7832 & 2009 Feb 27 & 4890.26684 & 7.39 & 30 & 17.7$^\dag$  \\
S7834 & 2009 Feb 28 & 4891.25647 & 4.29 & 30 & 17.7$^\dag$  \\
S7840 & 2009 Mar 03 & 4894.30579 & 1.15 & 30 & 17.7$^\dag$  \\
S7842 & 2009 Mar 18 & 4909.24058 & 5.42 & 30 & 17.9$^\dag$  \\
S7843 & 2009 Mar 19 & 4910.23990 & 5.51 & 30 & 17.9$^\dag$  \\
S7856 & 2009 Mar 23 & 4914.29437 & 1.68 & 30 & 17.9$^\dag$  \\[5pt]
\end{tabular}
{\footnotesize
\newline
$^\dag$ Out of eclipse. \hfill}
\label{warnertab1}
\end{table}
   
\subsection{Light curves on the orbital time scale}

\begin{figure}
\centerline{\hbox{\psfig{figure=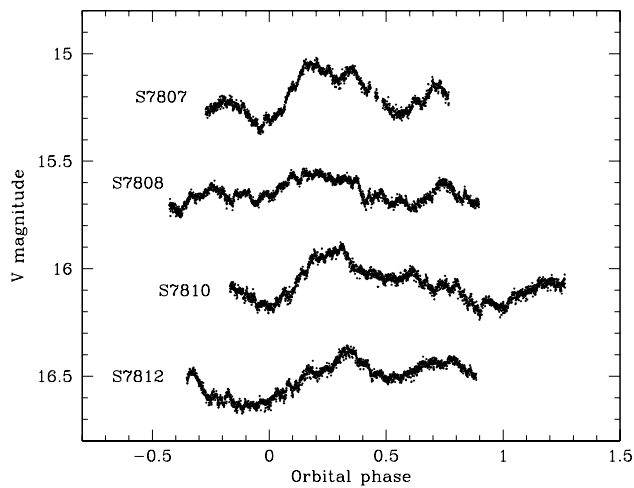,width=8.4cm}}}
  \caption{The 2008 light curves of V597 Pup phased on the orbital ephemeris
given in Eq.~\ref{warnereq1}. The light curves of runs S7808, S7810, S7812 have
been displaced by 0.4, 0.8 and 1.2 mag, respectively, for display purposes.}
 \label{warnerfig2}
\end{figure}

\begin{figure}
\centerline{\hbox{\psfig{figure=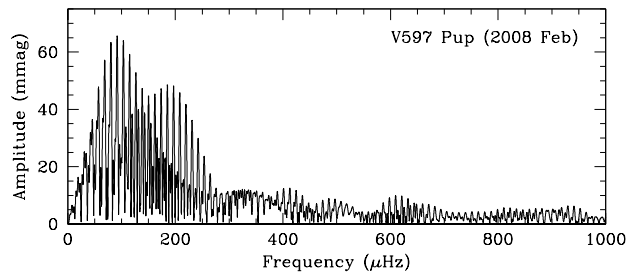,width=8.4cm}}}
  \caption{The low frequency Fourier transform of the combined
2008 observations of V597 Pup.}
 \label{warnerfig3}
\end{figure}

The light curves from the 2008 data are shown in Fig.~\ref{warnerfig2}, 
aligned in phase with the orbital period that we deduce below. A repetitive 
hump with an amplitude $\sim$ 0.2 mag is seen, but no obvious eclipse 
features; unfortunately our light curves had lengths not much greater than 
the (then unknown) orbital period. The Fourier transform (FT) of the 
combined light curves is shown in Fig.~\ref{warnerfig3}. The highest peak 
has a frequency of 91.8 $\mu$Hz, but there are aliases of almost equal 
amplitude either side at 80.3 and 103.3 $\mu$Hz (periods of 3.46, 3.03 
and 2.69 h). There is also a set of frequencies which could be first 
harmonics, at 173.7, 185.2, 196.7 and 208.1 $\mu$Hz. The uncertainties 
on these figures are $\sim$ 0.5 $\mu$Hz. 

\begin{figure}
\centerline{\hbox{\psfig{figure=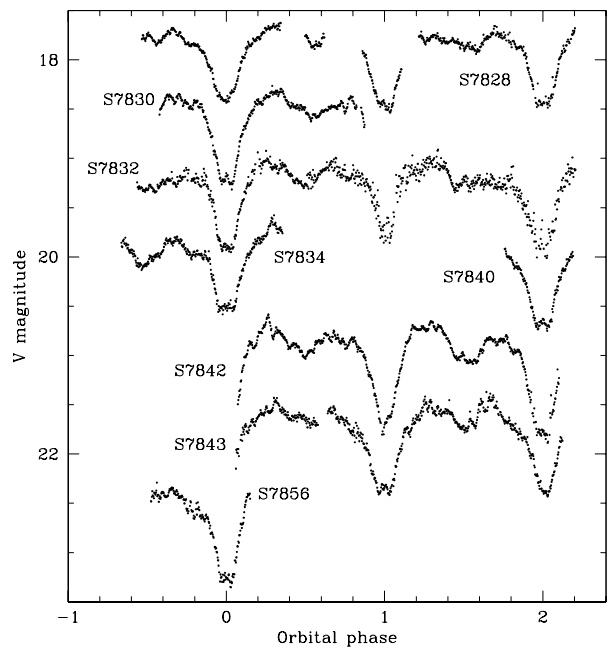,width=8.4cm}}}
  \caption{The 2009 light curves of V597 Pup phased on the orbital ephemeris
given in Eq.~\ref{warnereq1}. The light curve of run S7828 is displayed at the 
correct magnitude, the others have been displaced vertically for display 
purposes only.}
 \label{warnerfig4}
\end{figure}

Light curves from the 2009 observations are shown in Fig.~\ref{warnerfig4}, 
where it is clear that the drop in system brightness by $\sim$ 2.5 mag over one year 
has revealed broad eclipses $\sim$ 0.6 mag deep with respect to the 
out-of-eclipse level away from the reflection region (see below). The 
fundamental and harmonics in the FT of these light curves provide an 
orbital period of 2.6687 h (104.089 $\mu$Hz), which shows that the orbital 
modulation, with first harmonic, was present in the 2008 light curves. 
The ephemeris for minimum light is

\begin{equation}
{\rm HJD_{min}} =  2454888.3280 + 0\fd11119 \, {\rm E}.
\label{warnereq1}
\end{equation}

If the drop in brightness (by a factor $\sim$ 10) between the two sets 
of observations was due largely or entirely to the dispersal and cooling 
of nova ejecta far from the central binary, the eclipses and their related 
humps could well have been present even in the earlier light curves -- there 
are some features that resemble such diluted structures.

\begin{figure}
\centerline{\hbox{\psfig{figure=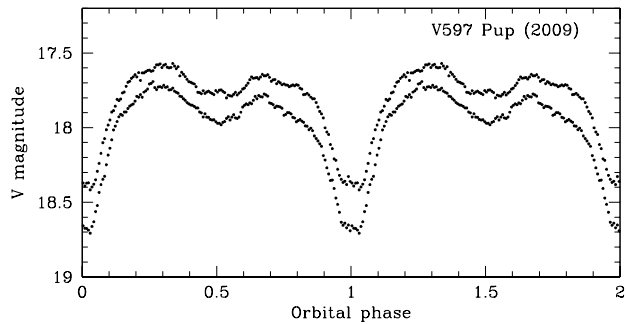,width=8.4cm}}}
  \caption{The average light curves of V597 Pup for the
February and March 2009 observations, plotted separately.}
 \label{warnerfig5}
\end{figure}

\begin{figure}
\centerline{\hbox{\psfig{figure=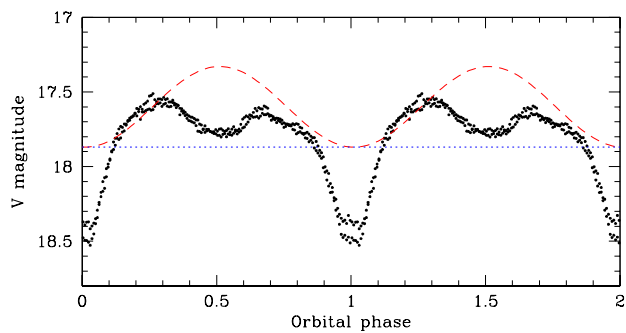,width=8.4cm}}}
  \caption{The average light curves of V597 Pup as in Fig.~\ref{warnerfig5} - the 2009 March light
curve is shifted vertically by 0.18 mag to match the 2009 February average magnitude - 
with overplotted the expected sinusoidal variation of a strong
reflection effect (dashed line) of $\sim$ 0.54 mag peak-to-peak amplitude. 
The depth of the deepest secondary eclipse (S7842 -- see Fig.~\ref{warnerfig4}) 
is an indication of the approximate minimum brightness at secondary eclipse, marked by the
horizontal dotted line.}
 \label{warnerfig6}
\end{figure}

The 2009 light curves for February and March 
averaged over the binary period are seen in 
Fig.~\ref{warnerfig5} and show repetitive structures between the primary 
eclipses. Remembering that these are of a fast nova only 15 months 
after outburst we would perhaps expect the white dwarf primary still to be 
very hot, so that at the small distance from the secondary implied by the 
short orbital period a strong reflection effect could be present. Between 
eclipses there is indeed what may be described as a reflection effect which 
would be 0.3 or 0.4 mag in amplitude were it not suppressed either side 
of phase 0.5 by a secondary eclipse of total width equal to that of 
primary eclipse. In Fig.~\ref{warnerfig6} we sketch what we suggest would 
be the reflection effect without any secondary eclipse; the asymmetry, 
namely the early fall in brightness before the start of primary eclipse 
is probably caused by partial obscuration of the bright face of the 
companion star by the gas stream passing from it to the outer edge of 
the disc. The reflection effect and secondary eclipse are similar, but 
of greater amplitude, to what we saw in the nova remnant DD Cir 
(Woudt \& Warner 2003), where we ascribed a shallow secondary eclipse 
to an optically thick almost edge-on disc passing across the heated 
hemisphere of the companion star. In V597 Pup the total widths of the 
primary and secondary eclipses are both $\sim$ $\pm$0.15 of $P_{orb}$.

There is great variability in the profile and depth of the secondary eclipse,
as could be expected from a disc of rapidly changing structure. Figs.~\ref{warnerfig5}
and \ref{warnerfig6} show that the 0.2 mag decrease in brightness from February to March 2009 has 
resulted in a deeper primary eclipse as the fading ejecta reduce the amount
of in-fill.

Inspection of the individual light curves in Fig.~\ref{warnerfig4} shows 
that during primary and secondary eclipses there is flickering similar 
to that seen out of eclipse, so at least one flickering source is still 
visible during eclipse. Apart from that, certainly the secondary and 
possibly the primary eclipse, appear flat bottomed, as in a total 
eclipse or a transit.

\subsection{Short period optical oscillations}

\begin{figure}
\centerline{\hbox{\psfig{figure=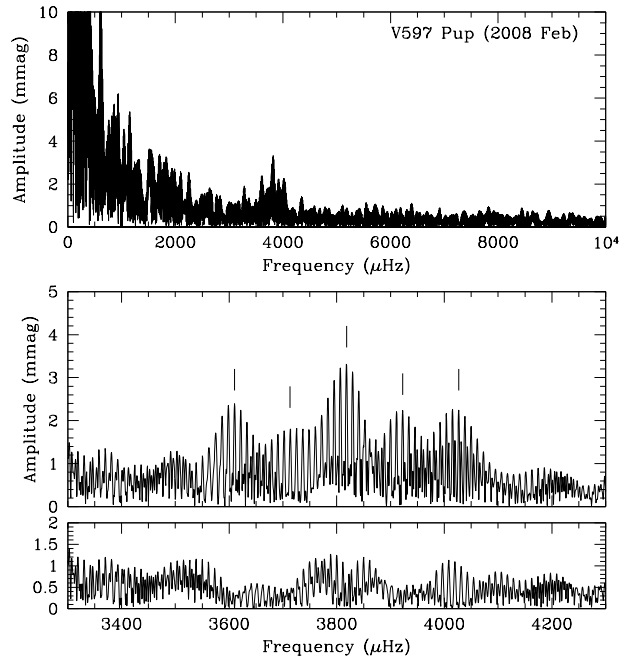,width=8.4cm}}}
  \caption{The high frequency Fourier transform of the combined
2008 observations of V597 Pup (top panel). The middle panel shows the 
detailed structure of the quintuplet around 3818 $\mu$Hz, where
the vertical bars correspond to the frequencies listed in 
Tab.~\ref{warnertab2}. The bottom panel shows the FT around 3818 $\mu$Hz
after pre-whitening (with a simultaneous 5 sinusoid least squares
fit) with the frequencies listed in Tab.~\ref{warnertab2}.}
 \label{warnerfig7}
\end{figure}

The FT of the combined 2008 light curves is shown in Fig.~\ref{warnerfig7}. 
There is red noise at low frequencies, associated with flickering and 
with harmonics of the orbital modulation, but an isolated
cluster of peaks in the vicinity of 3800 $\mu$Hz is evident. On detailed 
examination five components are found, with the frequencies and amplitudes 
listed in Tab.~\ref{warnertab2}, found from a 5 sinusoid simultaneous 
least squares fit. Uncertainties of the frequencies are all $\sim$ $\pm$0.5 
$\mu$Hz and of the amplitudes are $\pm$0.7 mmag. The phases (with 
arbitrary zero point) are quoted as fractions of cycles. The largest
amplitude signal is at 261.9 s. 

\begin{figure}
\centerline{\hbox{\psfig{figure=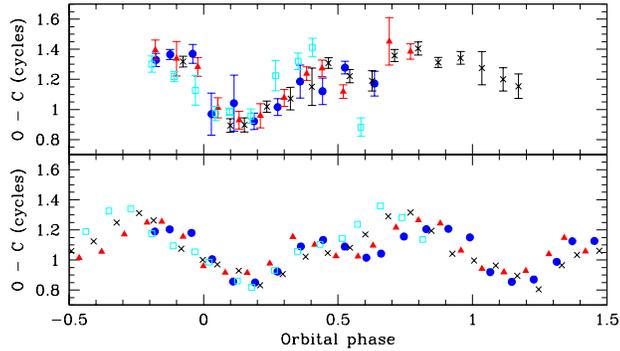,width=8.4cm}}}
  \caption{Phase variations (observed -- calculated) of the 3818 $\mu$Hz
signal in the 2008 data (observed: top panel; simulated (see text
for details): bottom panel). Different symbols represent data from different
nights. Each dot represents $\sim$ 6 cycles of the 3818 $\mu$Hz modulation, 
with a 50\% overlap.}
 \label{warnerfig7b}
\end{figure}

As further evidence for the significance of these signals we produce in Fig.~\ref{warnerfig7b}
an O -- C phase diagram relative to the central frequency and comparison with the same diagram
produced by a simulation using the five sinusoidal signals listed in Tab.~\ref{warnertab2}.

\begin{table}
 \centering
  \caption{The high frequencies (first harmonics) in V597 Pup.}
  \begin{tabular}{@{}ccccc@{}}
\multicolumn{3}{c}{\bf 2008} & \multicolumn{2}{c}{\bf 2009 February} \\
Frequency    &  Ampl. & Phase & Frequency  & Ampl. \\
($\mu$Hz)    &  (mmag)    &       & ($\mu$Hz) & (mmag) \\[10pt]
3610.0       &  2.2       & $-0.05 \pm 0.09$ &     &    \\
3713.4       &  2.0       & $-0.37 \pm 0.10$ & 3714.6 & 2.7 \\
3818.2       &  3.4       & $-0.21 \pm 0.06$ & 3817.8 & 6.3 \\
3922.4       &  1.6       & $-0.45 \pm 0.12$ &     &    \\
4026.8       &  2.3       & $+0.24 \pm 0.08$ & 4027.3 & 3.1 \\
\end{tabular}
\label{warnertab2}
\end{table}

The mean splitting between the components in 2008 is 104.15 $\mu$Hz, which, 
within errors, is the same as the orbital frequency. It is therefore clear 
that in 2008 a quintuplet of frequencies was present with internal splitting 
equal to the orbital frequency $\Omega$, which is the characteristic signature 
of an intermediate polar (IP) where the white dwarf primary's frequency $\omega$, 
and/or its reprocessed signal $\omega - \Omega$, acquires orbital sidebands. 
This occurs through reprocessing of a rotating beam of high energy radiation, 
emitted from the accretion zone(s) on the primary, off regions of varying 
cross section or varying visibility rotating in orbit (Warner 1986). This 
shows that the central binary was free from obscuration during the 2008 
observations and therefore that much of the luminosity came from distant 
ejecta and not an optically thick wind close to the primary. The apparent 
amplitude of the oscillations should be increased by a factor of at least 
10 to remove the diluting effect of these bright ejecta, so relative to 
the central binary luminosity the amplitudes are $\sim$ 0.03 mag. There is no 
significant extra power at the harmonics or subharmonic of the 261.9 s 
signal.

\begin{figure}
\centerline{\hbox{\psfig{figure=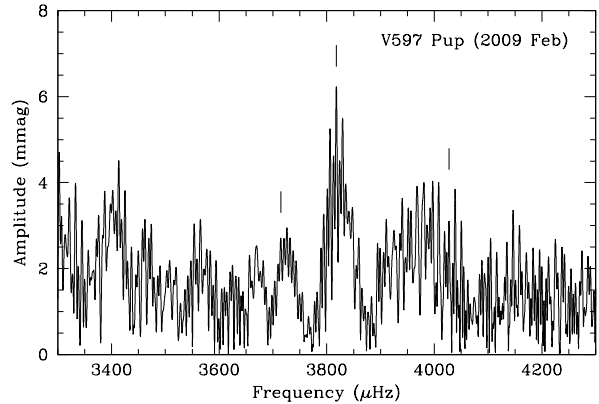,width=8.4cm}}}
  \caption{The high frequency Fourier transform of the combined
2009 February observations of V597 Pup around 3818 $\mu$Hz. 
The vertical bars correspond to the frequencies listed in 
Tab.~\ref{warnertab2}.}
 \label{warnerfig8}
\end{figure}

\begin{figure}
\centerline{\hbox{\psfig{figure=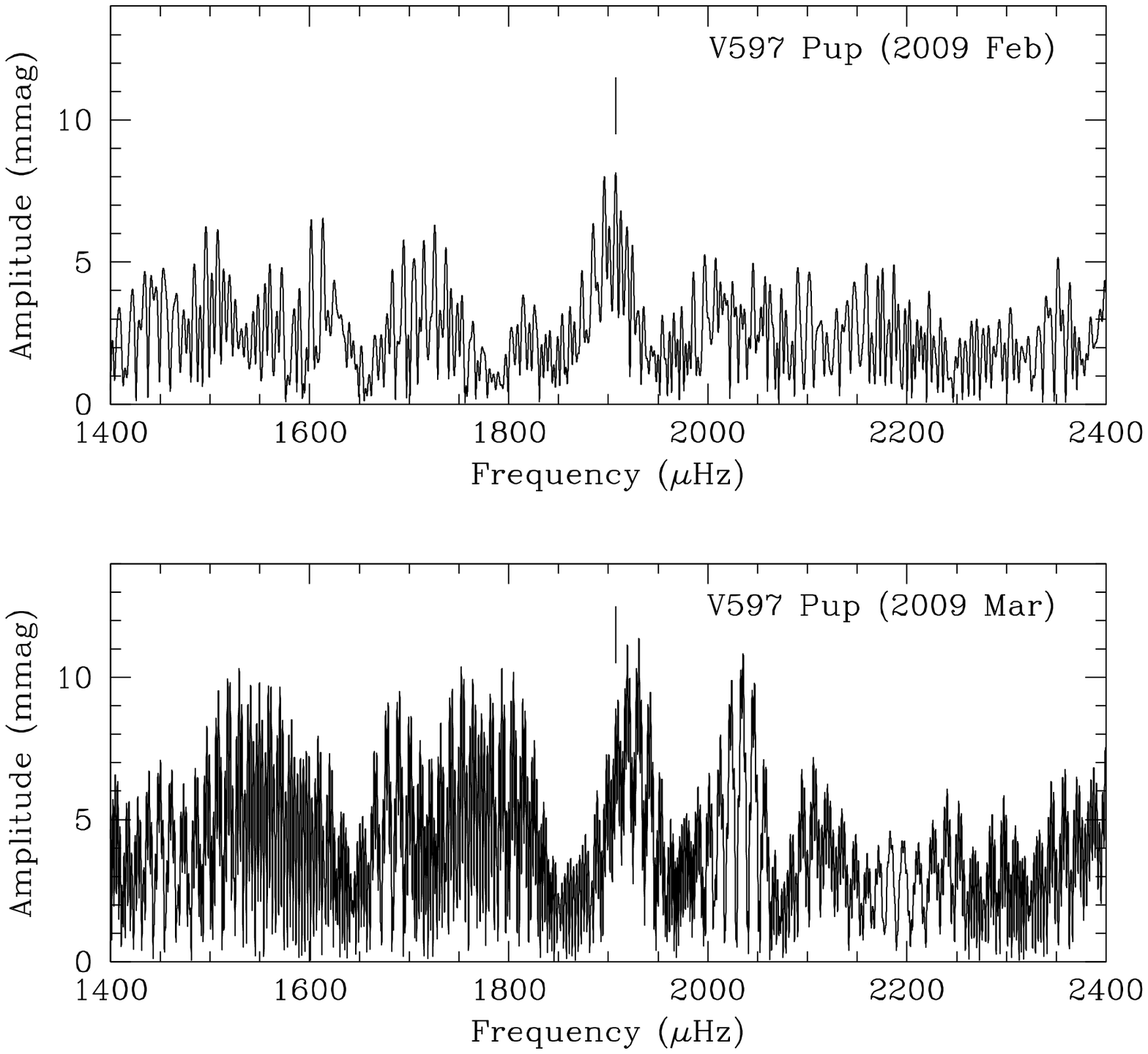,width=8.4cm}}}
  \caption{The high frequency Fourier transform of V597 Pup
around 1905 $\mu$Hz of the combined 2009 February observations 
(top panel), and the combined 2009 March observations (lower panel). 
The vertical bars correspond to the selected frequencies listed in 
Tab.~\ref{warnertab3}.}
 \label{warnerfig9}
\end{figure}

In the February 2009 light curves the sidebands are largely conflated 
with noise in the FT, but the central modulation is very clear -- see 
Fig.~\ref{warnerfig8} -- and near numerical coincidences with the 3713.4 
and 4026.8 $\mu$Hz sidebands are present. The amplitude, relative to the 
brightness of the central binary, has dropped by a factor $\sim$ 5. 
The average of the two years' frequency determinations is 3818.0 $\mu$Hz, 
or a period of 261.9 s. In the March 2009 light curves no components of 
the 2008 quintuplet are definitely present, but both February and March 
2009 light curves possess signals near `subharmonics' of the quintuplet: 
see Fig.~\ref{warnerfig9}. These are listed in Tab.~\ref{warnertab3}. 
In February 2009 the strongest signal was at 1907.5 $\mu$Hz (2 $\times$ 
262.1 s), with a nearby 1-d alias; in March 2009 there was a large 
amplitude window pattern which included the 1907.5 $\mu$Hz modulation 
as one of the aliases. The clear presence of a subharmonic to the 261.9 s 
signal shows that the periods listed in Tab.~\ref{warnertab2} are almost 
certainly first harmonics, not fundamentals. 

\begin{table}
 \centering
  \caption{The high frequencies (fundamentals) in V597 Pup.}
  \begin{tabular}{@{}cccc@{}}
\multicolumn{2}{c}{\bf 2009 February} & \multicolumn{2}{c}{\bf 2009 March} \\
Frequency$^\dag$    &  Ampl.  & Frequency$^\dag$  & Ampl. \\
($\mu$Hz)    &  (mmag) & ($\mu$Hz) & (mmag) \\[10pt]
             &         & {\it 1930.8 $\pm$ 0.3} & {\it 11.3 $\pm$ 2.8} \\
             &         & {\it 1919.2 $\pm$ 0.3} & {\it 11.1 $\pm$ 2.8} \\
             &         & {\it 1909.9 $\pm$ 0.3} & {\it  9.2 $\pm$ 2.8} \\
1907.5 $\pm$ 0.5        & 8.1 $\pm$ 0.5        & {    1907.5 $\pm$ 0.6} & {     8.9 $\pm$ 2.0} \\
{\it 1896.1 $\pm$ 0.5}  & {\it 8.0 $\pm$ 0.5}  &                        &                      \\[5pt]
\end{tabular}
{\footnotesize
\newline
$^\dag$ Aliases are given in italics. \hfill}
\label{warnertab3}
\end{table}

As is usual with IPs (Warner 2004) we found no evidence for dwarf nova 
oscillations in the FTs of V397 Pup.

\section{The structure of V597 Pup}

The drop in luminosity from 2008 to 2009 revealed an IP in the orbital 
period gap (Warner 1995), with a rotation period of 8.7 min. The 
reduction in amplitudes of the spin modulations over the year is 
probably the result of diminishing $\dot{M}$ as the primary cools and lessens 
its irradiation of the secondary.

It is not possible, with the flickering and variations from orbit 
to orbit, to give precise values for eclipse ingress and egress timings, but 
we can show the effect in the light curve of the various components. 
First, a simple calculation shows that the measured widths of the 
eclipses imply an unusually large disc radius. From the expected 
high mass for a very fast nova, and the mass and radii relationships 
for CV secondaries (Warner 1995), we adopt $M(1)$ = 1.3 M$_{\odot}$, 
$M(2)$ = 0.22 M$_{\odot}$, $R(2)$ = $1.9 \times 10^{10}$ cm. These lead 
to a separation $a = 7.82 \times 10^{10}$ cm, a radius $R(L)$ of the 
Roche lobe of the primary of $4.2 \times 10^{10}$ cm, and expected 
radius of a high $\dot{M}$ disc of $R_d = 0.7 R(L) = 2.9 \times 10^{10}$ 
cm. The first and last contact phases $\pm \phi$ of eclipses, relative to 
phase 0, assuming an inclination near 90$^{\circ}$, is found from 
$\sin \phi = \frac{[R(2) + R]}{a}$, where $R$ is selected from above. 
For the usual $R = R_d$ we find $\phi$ = 0.10, but for $R = R(L)$ we 
have $\phi$ = 0.14, which is close to the observed $\phi \sim 0.15$. 
The disc in V597 Pup fills the primary's Roche lobe to its maximal 
possible extent. The radius of the disc is considerably larger than 
that of the secondary star, which results in the flat bottomed secondary 
eclipse.

The phases of first and last contacts in the secondary eclipse, 
as the secondary star (made visible by irradiation from the hot 
primary) is eclipsed by the disc, relative to phase 0.5, are 
also $\pm \phi$, and the phases between second and third contact 
are $\pm (\phi - \theta)$, where $\sin \theta = \frac{[R(L) - R(2)]}{a}$, 
which are $\pm$0.09, in reasonable agreement with what is observed 
for the emergence of the secondary, but immergence takes longer 
than predicted, possibly because at that phase the illuminated gas 
stream is also being eclipsed by the disc.

The general agreement between observed and expected time scales of the 
secondary eclipses adds weight to our interpretation of this 
unusual light curve.

With the short orbital period, and mass ratio $M_2/M_1 = 0.22$ adopted 
here, we might have expected an elliptical disc and superhumps to be 
present (see Chapter 3 of Warner 1995), but there is no evidence for 
them in the light curves and FTs. The reason may be that with 
an accretion disc of such large radius, greatly exceeding the 3:1 
resonance radius that excites ellipticity, the resonance is ineffectual: 
Osaki \& Meyer (2003) have shown how in a disc of large radius the 2:1 
Lindblad resonance can suppress the 3:1 resonance. Our estimate of 
$\phi \sim 0.15$ in fact puts the outer edge of the accretion disc 
very close to the 2:1 resonance radius.

\section{V597 Pup as an intermediate polar}

Mukai (2008) lists over 30 bona fide IPs, among which there are none 
with relatively rapid rotations that possess deep eclipses suitable 
to provide detailed information about the structure of an IP. Eclipses 
are seen in XY Ari but it is an X-Ray source, hidden behind a dust cloud; 
DQ Her is far more favourable for observation but is not ideal as it has 
an inclination so high that much of the central region of the disc is 
hidden from view; DD Cen (Woudt \& Warner 2003) is too faint to study 
fully. Thus V597 Pup is the first relatively standard IP found to have 
eclipses -- which can be expected to grow a little deeper over the 
next few years as the residual emission from the ejecta decays. We analyse 
our observations further in order to develop a model for the system.

\subsection{Interpretation of the Quintuplet in the 2008 Light Curves}

The standard interpretation of multiple sidebands in IPs (Warner 1986) 
ascribes the (often very strong) $\omega - \Omega$ component to reprocessing 
of $\omega$ from the secondary star or the disc thickening near the bright 
spot or the inner Lagrangian point (where in such a large disc the
vertical gravity is very low)\footnote{The reader is reminded that 
sinusoidal modulation of $\omega$ by $\Omega$ generates signals at 
$\omega$ and $\omega \pm \Omega$.}, all of 
which may vary in cross section at frequencies $\Omega$ and 2$\Omega$, and can 
thus generate $\omega - 2\Omega$, $\omega$ and $\omega + \Omega$ signals, 
as well as $\omega - \Omega$.  This gives only four components, a quadruplet, 
whereas in V597 Pup we uniquely see a quintuplet. We suggest that the clue 
to understanding this comes from the unusual presence of a secondary 
eclipse: the $\omega - \Omega$ component is therefore modulated not 
only at $\Omega$ (with maximum at superior conjunction and minimum at 
inferior conjunction of the secondary), but also at 2$\Omega$ (another 
minimum being imposed by the secondary eclipse, where otherwise a 
maximum of reprocessing would occur). This generates $\omega - \Omega \pm 2\Omega$ 
components, so together with the standard quadruplet, the complete set will 
be $\omega - 3\Omega$, $\omega - 2\Omega$, $\omega - \Omega$, $\omega$, 
$\omega + \Omega$. From what we deduced in Section 2.2 the 2008 quintuplets 
must be the first harmonics of then unobservable fundamentals. 
   
This interpretation attributes the highest amplitude component, at 3818.0 $\mu$Hz, 
to 2($\omega - \Omega$), not to the harmonic of the primary's spin, 2$\omega$. 
Interestingly, the well known 71 s modulation in the nova remnant DQ Her, 
which is also a high inclination binary, has recently been re-interpreted 
(Saito \& Baptista 2009) as ($\omega - \Omega$) and not $\omega$, being a 
result of reprocessing from the thickened disc region around the bright 
spot.

\subsection{The Phase Variation of the 262 s modulation through Primary Eclipse}

The first phase shift discovered in the eclipse of an IP was for DQ Her 
(Warner et al.~1972), studied in great detail by Patterson, Robinson \& 
Nather (1978). The phase (observed -- calculated) of the 71 s modulation 
increases by $\sim 90^{\circ}$ from the beginning to mid-point of eclipse, 
jumps through $\sim 180^{\circ}$ and recovers from mid- to end of eclipse. 
By contrast, the stable Dwarf Nova Oscillations (DNOs) in the nova-like 
variable UX UMa decrease though $360^{\circ}$ during eclipse 
(Nather \& Robinson 1974); the same behaviour has been seen in HT Cas 
(for a review of eclipse phase changes see Chapter 8 of Warner 1995).

The phase changes have been successfully modeled by the reprocessing 
of high energy radiation emitted from accretion zones on the rotating 
primary (Chester 1979; Petterson 1980), and are a function of the relative 
amounts of direct radiation from the primary and those from the front and 
back portions of the accretion disc.

\begin{figure}
\centerline{\hbox{\psfig{figure=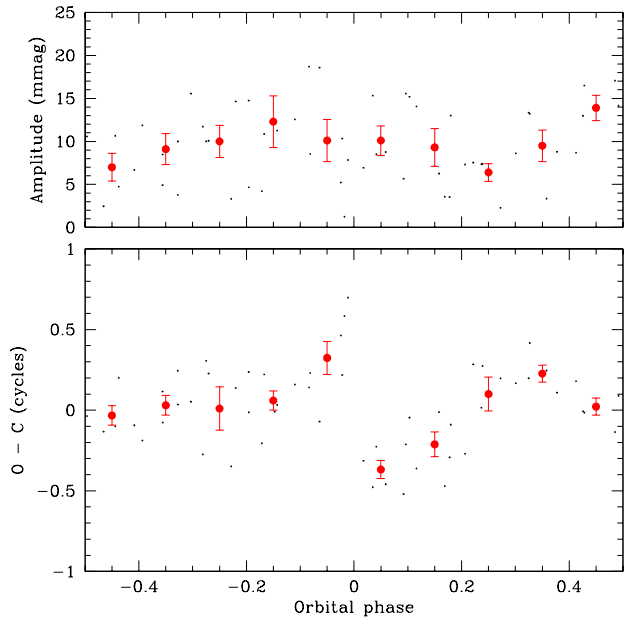,width=8.4cm}}}
  \caption{The amplitude (upper panel) and phase variation (observed -- calculated, 
lower panel) of the 3818 $\mu$Hz modulation as a function of 
orbital phase (Eq.~\ref{warnereq1}).}
 \label{warnerfig10}
\end{figure}

In V597 Pup the phase behaviour of the 262 s signal in 2009 is of a kind
that has not been observed before. 
Our initial impression was of a $360^{\circ}$ increase in phase during 
eclipse -- which might be thought to indicate that the primary rotates 
retrogradely in the system. But further investigation showed that 
Chester (1979) had anticipated the situation -- he pointed out that 
if the phase of the first harmonic is measured in a high inclination 
system (where only the beam sweeping across the back surface of the disc 
is observable, as in DQ Her) then a $180^{\circ}$ increase would be seen, 
followed by an instantaneous decrease of $360^{\circ}$ and a recovery 
increase of $180^{\circ}$. There is of course an ambiguity in how to 
plot such observed phase shifts, but we have chosen to use Chester's 
insight rather than claim a retrogradely rotating primary. The phase 
variations through eclipse are present in all the high quality light 
curves; these and their average are shown in Fig.~\ref{warnerfig10}.

There is no significant reduction in amplitude near phase 0, which may
be the result of the $\omega - \Omega$ signal having at least three
sources (secondary and two regions of disc thickening). Furthermore,
unlike DQ Her, where the radii of the secondary and disc are almost
equal (Petterson 1980), in V597 Pup the secondary is considerably 
smaller and leaves much of the disc uncovered at phase 0.

As seen in Fig.~\ref{warnerfig10}, there are no significant
phase changes associated with the secondary eclipse.

\subsection{The Phase Variation of the 524 s modulation}

\begin{figure}
\centerline{\hbox{\psfig{figure=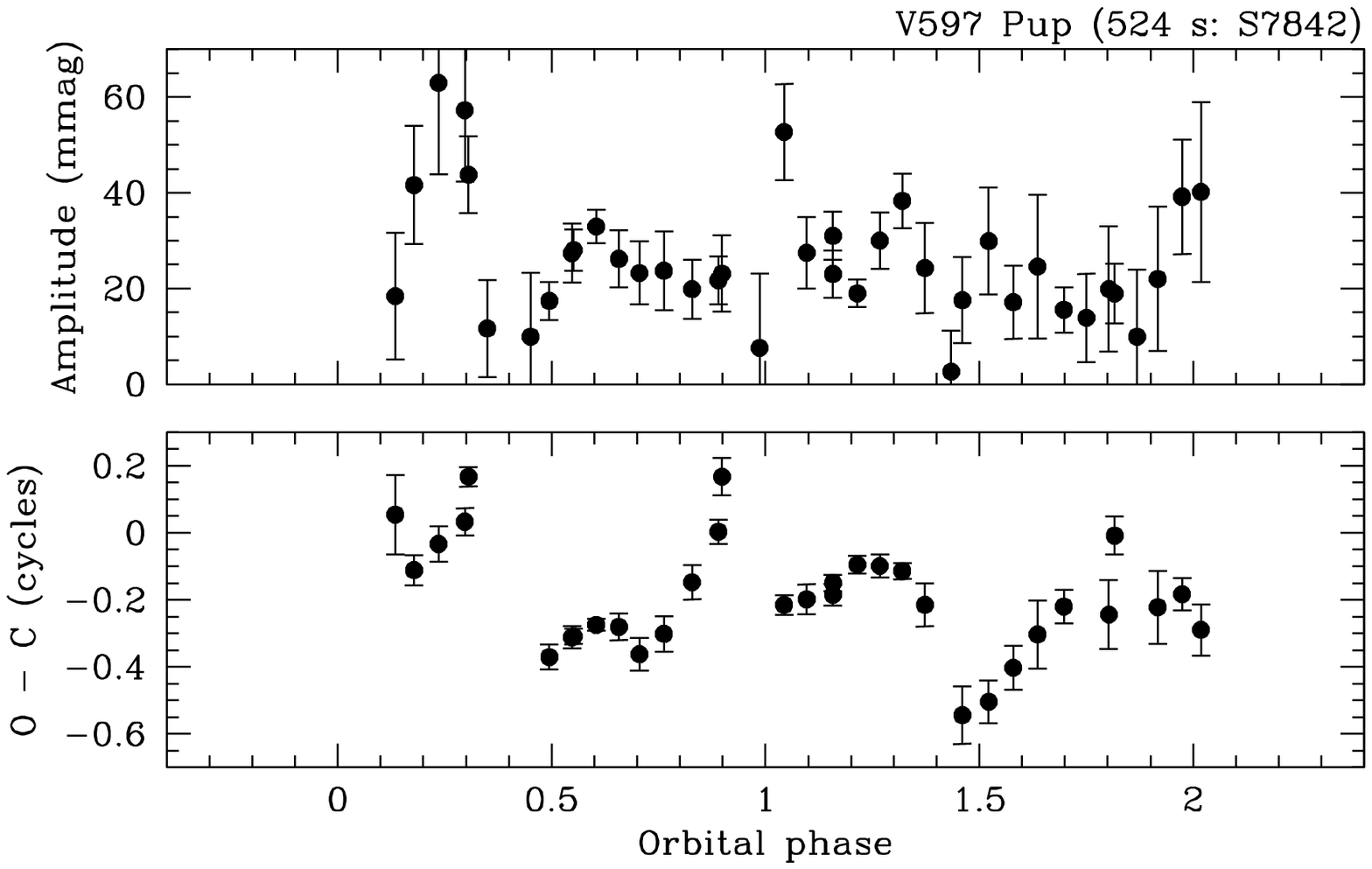,width=8.4cm}}}
\centerline{\hbox{\psfig{figure=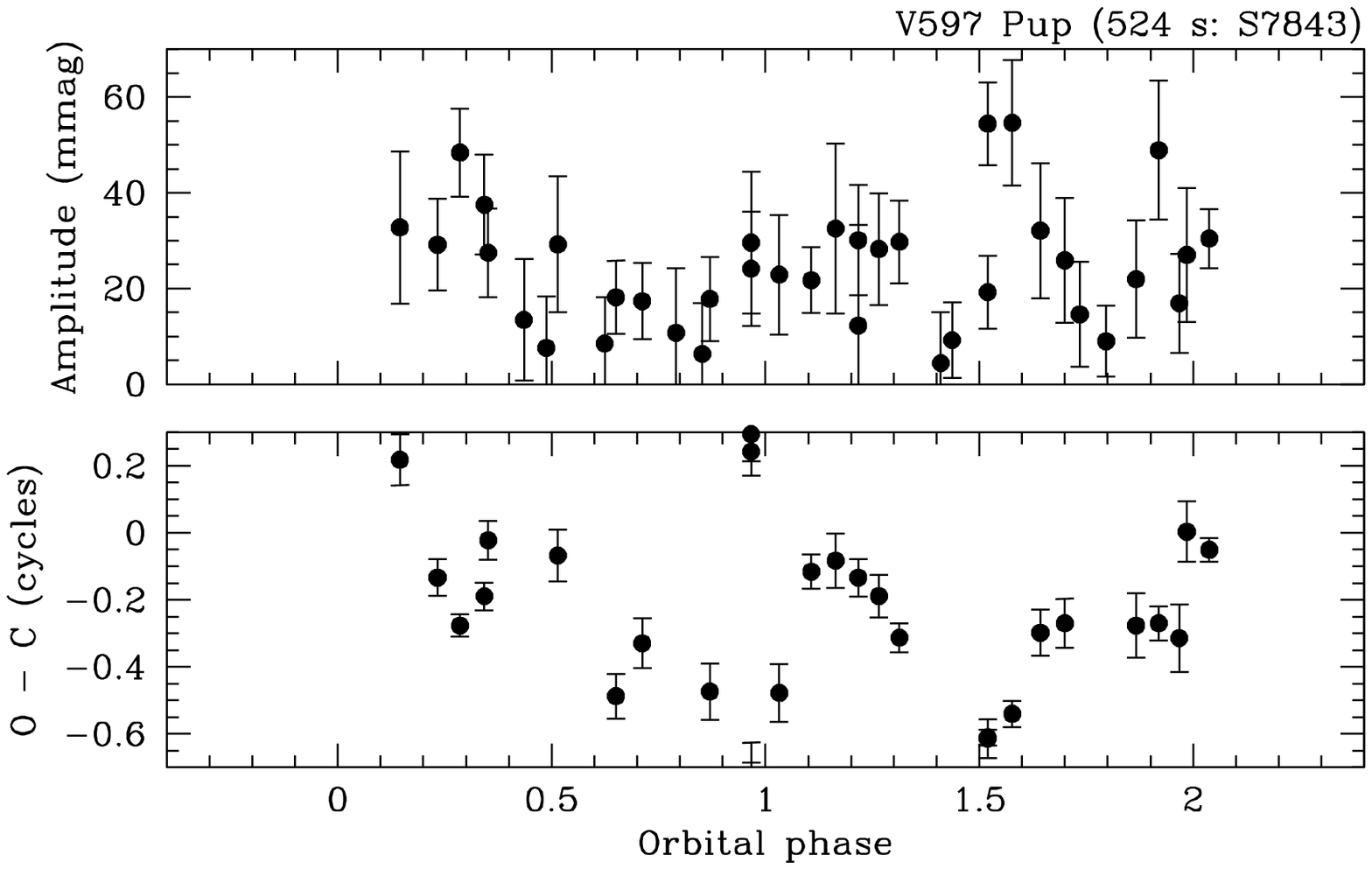,width=8.4cm}}}
  \caption{The amplitude and phase variation (upper two panels, respectively) of the
524 s modulation in run S7842. The lower two panels show the amplitude and phase
variation of the 524 s modulation in run S7843. Each dot represents $\sim$ 2 cycles 
of the 524 s modulation, with a 50\% overlap. In the phase diagrams, only those points are
plotted where the amplitude of the modulation is larger than 10 mmag.}
 \label{warnerfig11}
\end{figure}

The phase and amplitude variations of the 1907.5 $\mu$Hz (524 s) fundamental
modulation seen in our best light curves (runs S7842 and S7843 of March 2009, both
of which cover two orbits) are shown in Fig.~\ref{warnerfig11}. The results 
through the eclipses are
not clear cut, which may result from varying contributions of different sources
to the $\omega - \Omega$ signal, but is largely due to the modulation period being
comparable to the ingress and egress durations. 

However, there is a DQ Her-like variation
of phase centred on the first primary eclipse of S7842, not repeated in the second
eclipse, and a suggestive rise in the second eclipse of S7843. In addition there is
a reverse phase change (also $\sim 180^{\circ}$) in the secondary eclipses of S7842 and a 
probably similar effect in the second secondary eclipse of S7843 (in the first
of the secondary eclipses of this run the amplitude of the signal is quite low).
There are also some indications of lower amplitudes at the times of both primary
and secondary eclipses.

\section{Conclusion}

V597 Pup is unique in several ways and will repay further 
photometric and spectroscopic investigation before it fades another 
two magnitudes to its pre-eruption brightness. It is the first 
cataclysmic variable found to have a deep secondary eclipse -- 
the result of a background generated by the temporary extra brightness 
of one face of the otherwise low luminosity secondary star. Detailed study 
of this eclipse could reveal more about the structure of the outer parts 
of the accretion disc.

As an intermediate polar V597 Pup offers a rare opportunity to study 
reprocessing in a very high inclination system. It has some parallels 
with DQ Her, but the fundamental and first harmonics of the rotation 
and reprocessed signals are more complex.

\section*{Acknowledgments}

Our research is supported by the University of Cape Town and by 
the National Research Foundation.

\end{document}